\documentstyle[art12]{article}
\begin{document}
\newtheorem{thm}{Theorem}[section]
\newtheorem{prop}[thm]{Proposition}
\newtheorem{lem}[thm]{Lemma}
\newtheorem{dfn}[thm]{Definition}
\newtheorem{conj}[thm]{Conjecture}
\newtheorem{cor}[thm]{Corollary}
\newcommand{\bean}{\begin{eqnarray*}}
\newcommand{\eean}{\end{eqnarray*}}
\newcommand{\ed}{\end{document}}
\newcommand{\pr}{\prime}
\newcommand{\ppr}{\prime\prime}
\newcommand{\cE}{{\tilde E}}
\newcommand{\vphi}{{\varphi}}
\newcommand{\oO}{O(k^{-1})}
\newcommand{\be}{\begin{equation}}
\newcommand{\ee}{\end{equation}}
\newcommand{\barr}{\begin{array}}
\newcommand{\earr}{\end{array}}
\newcommand{\bea}{\begin{eqnarray}}
\newcommand{\eea}{\end{eqnarray}}
\newcommand{\pa}{\partial}
\newcommand{\xx}{\hbox{}^*_*}
\newcommand{\sds}{\subset\hskip - 1em +}
\newcommand{\qed}{\hfill \fbox{}\medskip}
\newcommand{\proof}{\medskip\noindent{\it Proof.}\quad }

\title{First class functions in constrained second class systems.}
\author{A.V.Bratchikov \\ Kuban State Technological University,\\ 2
Moskovskaya Street, Krasnodar, 350072, Russia\\
E-mail:bratchikov@kubstu.ru} \date {} \maketitle

\begin{abstract}
Generators of the algebra of
first class functions in a
system with second class constraints are found.
It is shown that first class functions form
algebras with respect to the Dirac bracket and pointwise
multiplication.The subspace of functions vanishing on constraint
surface are ideals of these algebras. The corresponding quotient
algebras are isomorphic to the algebras of
phase variables in the Dirac bracket formalism.
Explicite expressions for generators and brackets of the
algebras under consideration are obtained.
\end{abstract}
\bigskip




\section {Introduction}
First class functions  play an important role in covariant
description of constrained second class systems.
Hamilton equations of such
systems are defined
modulo the functions which vanish on constraint
surface.
In a recent article \cite {Br} it was shown that the
corresponding  quotient Dirac bracket algebra is isomorphic to a
quotient of the algebra of first class functions with respect to the
original Poisson bracket. In spite of the algebra of first class
functions was introduced long ago \cite {D}, its generators and
brackets were not described.

In the present article we find explicite expressions for generators
of this algebra
in a system with second class constraints and compute corresponding
Poisson brackets.

We show that first class functions form a
Dirac bracket algebra and compute the corresponding bracket.
The
first class functions which vanish on constraint surface form an ideal
of this algebra. This enables us to construct a new quotient algebra
with respect to the Dirac bracket. The new algebra is
isomorphic to the original quotient
Dirac bracket algebra. Isomorphic image
of constrained Hamilton equations can be treated as a partial
fixing of gauge invariance.

The algebra of functions on phase space
modulo the functions which vanish on constraint
surface is also an algebra (with respect to the  pointwise
multiplication).This enables us to obtain a new form of
constrained Hamilton equations.

We observe that
first class functions form an algebra with respect to the
pointwise multiplication.The same is true for
these functions modulo the functions which vanish on constraint
surface.We show that this quotient algebra is isomorphic to
the algebra of phase variables of the original
constrained Hamilton equations.

The Letter is organized as follows. In Section 2 we
review some properties of constrained Hamilton equations
and describe the algebraic connection between these equations and
Hamilton equations on quotient spaces.  In Section 3 we find an
explicite realization of first class functions.  The space of first
class functions and a quotient are studied as Poisson bracket
algebras.In Sections 4,5  these spaces are studied respectively as
Dirac bracket algebras and algebras with respect to the pointwise
multiplication.


\section {Constrained Hamilton equations}
Let $M$ be a phase space with the phase variables $\eta_n,\,n=1...2N,$
and the Poisson bracket $
[\eta_m,\eta_{n}
]=\omega_{mn}(\eta).
$
Let $H(\eta)$ be the original hamiltonian
and $\varphi_j(\eta), j=1...2J,$ the second class constraints
$det [\varphi_j,\varphi_{k} ]|_{\varphi=0} \ne 0.$

The dynamic of the system under consideration is described by the
Hamilton equations (see, e.g.,\cite {GT})
\bea
\label{h1}
\frac {d} {dt}
\eta_n = [\eta_n,H_T ],\qquad  \varphi_j=0.
\eea
Here
$H_T=H+{\lambda_j\varphi_j}$
and functions $\lambda_j=\lambda_j(\eta)$ are defined by
the  equation \bea \label  {hh12}
[H_T, \varphi_j]|_{\varphi=0}=0.
\eea

Using (\ref {hh12}) one can write equations
(\ref{h1})
as
\bea \label{h3}
\frac {d} {dt}
\eta_n = [\eta_n,H_T ]_D,
\qquad \varphi_j=0.
\eea
Here the Dirac bracket was introduced
\bea \label{Dbr}
[g,f]_D=[g,f]- [g,\varphi_{j}]c_{jk}[\varphi_{k},f],\qquad
c_{jk}[\varphi_{k},\varphi_{l}]=\delta_{jl} .
\eea

Let $A$ be the
space
of  functions on $M$
and $\Phi\subset A$ be the subspace of the functions
which vanish on constraint surface.It is known that $\Phi$ is
an ideal
and $A/\Phi$ is an algebra
with respect to the Dirac
bracket.

Let $F: A \to A/\Phi$ be the canonical homomorphism
\bean  \label {T}
F(g)=\{g\}
\eean
where
$\{g\} \in
A /\Phi $ is the coset represented by $g.$
The homomorphic image of equations
(\ref {h3}) in $A/\Phi$ is
\bea \label {dr}
\frac {d} {dt}
\{\eta_n\}=\{[\eta_n,H_T]_D \}.
\eea

Equations (\ref {dr}) were introduced by Dirac.In the original
notations \cite {D} they are written \bean \label{h35} \frac {d} {dt}
\eta_n \approx [\eta_n,H_T]_D, \eean where $f\approx g$ means that
$f-g\in \Phi.$

Equations (\ref {dr}) can be written as \cite {Br}
\bea \label{h37}
\frac {d} {dt}
\{\eta_n\}=[\{\eta_n\},\{H\})]_D.
\eea
To rewrite these equations in another form we need the
proposition:

\vspace{3mm}
\noindent
PROPOSITION 2.1.
{\it $\Phi$ is and
ideal of $A$ and $A/\Phi$ is an algebra with respect to
the pointwise multiplication.}

\proof The proof is straightforward.
\qed

\vspace{3mm}
\noindent
COROLLARY. In $A/\Phi$
\bean
f(\{\eta\})= \{f(\eta)\}.
\eean

\vspace{3mm}
From this it follows that
Hamilton equations (\ref{h37}) can be written as
\bea \label{h3788}
\frac {d} {dt}
\{\eta_n\}=[\{\eta_n\},H(\{\eta\}) ]_D
\eea
and
\bea \label{h37888}
\varphi_j(\{\eta\})=\{0\}.
\eea
Equations (\ref{h37888}) tell us that
cosets $\{\eta_n\},
n=1,\ldots,2N,$
represent the phase variables on constraint surface.

Dirac bracket algebra
$A /\Phi$
is isomorphic to Poisson bracket algebra
$\Omega/\Upsilon$ \cite {Br}.Here
$\Omega$ is the algebra of first class functions
and
$\Upsilon =
\Omega \cap \Phi
.$
The image of Hamilton
equations (\ref {h3788}) in  $\Omega/\Upsilon$
is
\bea \label{h777} \frac {d} {dt} \{\tilde \eta_n\}^\bullet= [\{\tilde
\eta_n \}^\bullet,\{H_T\}^\bullet].  \eea
Here
\bea \label {et}
{\tilde \eta}_n
=\eta_n - [\eta_n,\varphi_{j}]c_{jk}\varphi_{k}
\eea
and  $\{g\}^\bullet \in \Omega/\Upsilon$ denotes the class
represented by $g\in \Omega.$


\vspace{3mm}
\section {Poisson bracket algebras of first class functions}
To describe elements of  $\Omega$ explicitly let us consider the
equations
\bea
\label{Un} [\tilde g
,\varphi_i
]\in \Phi
\eea
 with the initial condition
\bea \label {ic1}
\tilde  g{
}\in \{g
\}.
\eea

It is easy to see that the function $
u_{ij}(\eta)
\varphi_{i}
\varphi_{j}$
satisfies (\ref {Un}).
Hence a solution to these equations
can be represented
in the form
\bea  \label{geg}
\tilde g =g + l_{i}\varphi_{i}
+u_{ij}
\varphi_{i}
\varphi_{j}
\eea
for some $l_{i},
u_{ij}\in A
$

Substituting (\ref{geg}) into  (\ref{Un}) we find
$l_i =-[g,\varphi_{i}]c_{ij}.$
Hence
\bea \label{geg3345}
\tilde g =L(g)+u
\eea
Here $L(g)=g-[g,\varphi_{i}]c_{ij}\varphi_{j}, u=
u_{ij}
\varphi_{i}
\varphi_{j}$ and
$u_{ij}(\eta)$ are arbitrary functions.

In particular for $\tilde v \in \{0\}$ we have
\bea \label{geg3346}
\tilde v =v_{ij}
\varphi_{i}
\varphi_{j}.
\eea
Here
$v_{ij}(\eta)$ are arbitrary functions.

Taking into account equations (\ref
{geg3345},\ref {geg3346}), we have the proposition:

\vspace{3mm}
\noindent
PROPOSITION 3.1.
{\it Algebras $\Omega,\Upsilon$ consist of all possible
expressions}
(\ref {geg3345}) {\it and} (\ref {geg3346})
{\it respectively, where $f,u_{ij}, v_{ij}\in A.$}

\vspace{3mm}
\noindent
Let
\bea  \label {fcf12}
\tilde g_a=
L(g_a) + u_a,\qquad  u_a \in \Upsilon,
\eea
$a=1,2,$ be first class functions
 and
\bean \label{geg3344}
l_{aj} =-[g_{a},\varphi_{i}]c_{ij}.
\eean


\vspace{3mm}
\noindent
PROPOSITION 3.2.
{\it
The Poisson bracket for
 first class functions} (\ref {fcf12})
{\it is given by
\bean \label {cr}
[\tilde g_1,\tilde g_2]=L([g_1,g_2]_D)+\tilde u_{12},
\eean
\bean
\tilde u_{12}=[l_{1i},l_{2j}]\varphi_{i}\varphi_{j}+
[L(g_1),u_2]+[u_1,L(g_2)]+[u_1,u_2]
\in \Upsilon.
\eean }

\proof   Straightforward calculation.
\qed


This  proposition  my  be  partly checked as follows.
It is easy  to see that $[\tilde g_1,\tilde g_2]\in \{[g_1,g_2]_D\}.$
On the other hand $[\tilde g_1,\tilde g_2]$ is a first class function
and hence it may  be written in the form (\ref {geg3345})
\bean \label{geg34}  [\tilde g_1,\tilde g_2] =L([g_1,g_2]_D) + v,\qquad
v\in \Upsilon. \eean

\vspace{3mm}
\noindent
COROLLARY
{\it For
$\{\tilde g_1\}^\bullet,\{\tilde
g_2\}^\bullet \in
\Omega/\Upsilon$ we have}
\bea
\label {gu112}
[\{\tilde g_1\}^\bullet,\{\tilde
g_2\}^\bullet]=\{L([g_1,g_2]_D)\}^\bullet.  \eea

\vspace{3mm}
\noindent
Let us
compare the functions
$\tilde g(\eta)$ (\ref {geg3345}), $g(\tilde \eta)$ and
$\tilde g(\tilde \eta)$
where
$\tilde \eta_n \in \Omega$ is given by (\ref {et}).
One can check that
$g(\tilde \eta)$ and
$\tilde g(\tilde \eta)$  as well as  $\tilde g(\eta)$ are first class
functions.The initial conditions for these functions read
$\tilde g(\eta), g(\tilde \eta),\tilde g(\tilde \eta)\in \{g(\eta)\}$
and hence \bean \label {mm}
\tilde g(\eta)=g(\tilde \eta)+u=\tilde g(\tilde \eta)+v
\eean for some
$u,v \in \Upsilon.$

From this it follows
\bea \label {msr}
\{\tilde g(\eta)\}^\bullet=\{g(\tilde \eta)\}^\bullet=
\{\tilde g(\tilde \eta)\}^\bullet.
\eea

Due to equation (\ref {hh12})  Hamiltonian $H_T$  is a first class
function.  Using  (\ref {msr}) we have \bea \label {mm12}
\{H_T(\eta\}^\bullet=\{H_T(\tilde \eta)\}^\bullet .
\eea

\vspace{3mm}
\noindent
\section {Dirac bracket algebras of first class functions}
PROPOSITION 4.1.

(i){\it  $\Omega$ is an algebra with respect to the Dirac bracket.}

(ii){\it $\Upsilon$ is and
ideal of $\Omega$ and $\Omega/\Upsilon$ is an algebra with respect to
the Dirac bracket.}

{\proof}For  $g,f \in \Omega$ the first term in the
r.h.s. of equation (\ref {Dbr}) is a first class function.
The second one is quadratic in constraints and hence is also a
first class function.This proves the first statement.

For $g\in \Omega$ and $f \in
\Upsilon$ we have $[g,f]_D\in \Upsilon$.Hence
$\Upsilon$ is an
ideal of $\Omega$ and $\Omega/\Upsilon$ is an algebra with respect to
the Dirac bracket.
\qed

\vspace{3mm}
\noindent
PROPOSITION 4.2.
{\it The Dirac bracket for first class functions} (\ref {fcf12})
{\it is given by }
\bean \label {w}
[\tilde g_1,\tilde g_2]_D= L([g_1,g_2]_D)+\tilde v_{12},
\eean
\bean
\tilde v_{12}=[l_{1i},l_{2j}]_D\varphi_i\varphi_j
+
[L(g_1),u_2]_D+[u_1,L(g_2)]_D+[u_1,u_2]_D
\in \Upsilon.
\eean
\proof   Straightforward calculation.
\qed

\vspace{3mm}
\noindent
COROLLARY.
{\it The Dirac bracket for
$\{\tilde g_1\}^\bullet,\{\tilde
g_2\}^\bullet \in
\Omega/\Upsilon$ is given by}
\bea
\label {gg112}
[\{\tilde g_1\}^\bullet,\{\tilde
g_2\}^\bullet]_D=\{L([g_1,g_2]_D)\}^\bullet.  \eea

\vspace{3mm}
\noindent
PROPOSITION 4.3. {\it Dirac bracket algebra  $\Omega/\Upsilon$ is
isomorphic to Poisson bracket algebra  $\Omega/\Upsilon.$}

{\proof} From equations  (\ref {gu112}) and   (\ref {gg112})
we have
\bea  \label {www}
[\{\tilde g_1\}^\bullet,\{\tilde g_2\}^\bullet]= [\{\tilde
g_1\}^\bullet\{\tilde g_2\}^\bullet]_D.
\eea
This proves the statement.
\qed

By using the result of \cite {Br} one obtains the following corollary:

\vspace{3mm}
\noindent
COROLLARY.{\it  Dirac bracket algebras $\Omega/\Upsilon$  and
$A/\Phi$ are isomorphic.}

\vspace{3mm}
The realization of Hamilton equations (\ref {h37}) in
Dirac bracket algebra $\Omega/\Upsilon$ is
\bea
\label {sss} \frac {d} {dt} \{\tilde \eta_n\}^\bullet=
[\{\tilde \eta_n \}^\bullet,\{H_T\}^\bullet)]_D.
\eea
Due to (\ref {www}) equations (\ref {sss}) and  (\ref {h777})
are identical.

\section {Algebras of first class functions}
In this section we study the spaces under consideration
as algebras with respect to
the pointwise multiplication.

\vspace{3mm}
\noindent
PROPOSITION 5.1.

(i) {\it $\Omega$ is an algebra with respect to the
pointwise multiplication.}

(ii){\it $\Upsilon$ is and
ideal of $\Omega$ and $\Omega/\Upsilon$ is an algebra with respect to
the pointwise multiplication.}

\proof The proof is straightforward.
\qed

\vspace{3mm}
\noindent
COROLLARY 1 ( Leibniz rule ). {\it For
$\{f\}^\bullet,\{g\}^\bullet,\{h\}^\bullet \in \Omega/\Upsilon$ one
has} \bean
[\{f\}^\bullet,\{g\}^\bullet\{h\}^\bullet]=[\{f\}^\bullet,\{g\}^\bullet]
\{h\}^\bullet+\{g\}^\bullet[\{f\}^\bullet,\{h\}^\bullet].
\eean

\vspace{3mm}
\noindent
COROLLARY 2. {\it  In $\Omega/\Upsilon$
\bean
g(\{ \tilde \eta \}^\bullet)=\{ g (\tilde \eta) \}^\bullet.
\eean             }

Using COROLLARY 2 and equation (\ref {mm12}) one can rewrite
Hamilton equations  (\ref {h777}) in the
form
\bean \label{h79} \frac {d} {dt} \{\tilde \eta_n\}^\bullet=
[\{\tilde \eta_n \}^\bullet, H_T(\{\tilde \eta\}^\bullet)].  \eean

\vspace{3mm}
\noindent
PROPOSITION 5.2. {\it
For $\tilde g_1,\tilde g_2$} (\ref {fcf12}) {\it we have
\bean    \label {rrr}
\tilde g_1\tilde g_2=L(g_1g_2)+
w_{12},
\eean
\bean
w_{12}=
l_{1i}l_{2j}\varphi_i\varphi_j +\tilde g_1 u_{2} +u_{1}\tilde g_2\in
\Upsilon
.
\eean }

\proof   Straightforward calculation.
\qed

\vspace{3mm}
\noindent
COROLLARY.
{\it For $\{\tilde g_1\}^\bullet,\{\tilde g_2\}^\bullet \in
\Omega/\Upsilon$ we have \bean \{\tilde g_1\}^\bullet\{\tilde
g_2\}^\bullet=\{L(g_1g_2)\}^\bullet.  \eean}

\vspace{3mm}
\noindent
PROPOSITION 5.3.
{\it $\Omega/\Upsilon$ and $A/\Phi$ are
isomorphic with respect to the pointwise multiplication.}

\proof
Define the linear function $T:
\Omega /\Upsilon \to A /\Phi$
\bea  \label {T}
T(\{g\}^\bullet)=\{g\} .
\eea
This function has the inverse
and hence determines the one-to-one correspondence
between elements of
$\Omega /\Upsilon$ and $A /\Phi$ \cite {Br}.

Using the definitions of $\Omega /\Upsilon,A /\Phi$ and $T$
we have
\bean T(\{g\}^\bullet\{f\}^\bullet)=
T(\{gf\}^\bullet)= \{gf\}=\{g\}\{f\} =
T(\{g\}^\bullet)T(\{f\}^\bullet).  \eean
and hence $T$ is a
homomorphism.
This proves the statement.
\qed

In equation (\ref {T})
$\{g\}^\bullet \subset \{g\}.$ Hence transition from
$A/\Phi$ to $\Omega/\Upsilon$ can be treated as
a partial fixing of gauge invariance.


\bigskip

{\bf Acknowledgements}

The author thanks RFBR, grant 03-02-96521, for support.

\end{document}